\documentclass[journal]{IEEEtran}
\usepackage{bm}
\usepackage{ushort}
\usepackage{mathtools}
\usepackage{amsmath, amssymb}
\usepackage{trfsigns}
\usepackage[]{graphicx}
\usepackage{psfrag}
\usepackage[]{algorithm2e}

\ifCLASSINFOpdf
\else
\fi
\begin{document}
%
\title{A statistical model for the excitation of cavities through apertures}
%
%
%

\author{Gabriele~Gradoni,~\IEEEmembership{Member,~IEEE,}
        Thomas~M.~Antonsen,~\IEEEmembership{Fellow,~IEEE,}
        Steven~M.~Anlage,~\IEEEmembership{Member,~IEEE,}
        and~Edward~Ott,~\IEEEmembership{Life~Fellow,~IEEE}
\thanks{G. Gradoni, T. M. Antonsen, and E. Ott are with the Institute for Research in Electronics and Applied Physics, University of Maryland, College Park,
MD, 20742 USA e-mail:ggradoni@umd.edu.}
\thanks{S. Anlage is with the Center for Nanophysics and Advanced Materials, University of Maryland, College Park,
MD, 20742 USA}
\thanks{Manuscript received December, 2013; accepted November, 2014.}}

\maketitle

\begin{abstract}
In this paper, a statistical model for the coupling of electromagnetic radiation into enclosures through apertures is presented.
The model gives a unified picture bridging deterministic theories of aperture radiation, and statistical models necessary for capturing 
the properties of irregular shaped enclosures. 
A Monte Carlo technique based on random matrix theory is used to predict and study the power transmitted through the aperture into the enclosure. 
Universal behavior of the net power entering the aperture is found.
Results are of interest for predicting the coupling of external radiation through openings in irregular enclosures and reverberation chambers.
\end{abstract}


\IEEEpeerreviewmaketitle

\section{Introduction}\label{sec:intro}
\IEEEPARstart{T}{he} coupling of electromagnetic radiation into enclosures or cavities through apertures both electrically small \cite{Bethe_1944} and large \cite{Harrington} 
has attracted the interest of electromagnetic community for many years \cite{Harrington_1982}-\cite{Butler_1978}. 
Full solutions of this problem are particularly complicated because of the mathematical complexity in the solution of the boundary-value problem 
and because of the sensitivity of the solution to the detail of the enclosure's dimensions, content, and the frequency spectrum of the excitation.
These difficulties have motivated the formulation of a statistical description (known as the random coupling model, RCM \cite{zao1,zao2,Gradoni2014606}) of the excitation of cavities. 
The model predicts the properties of the linear relation between voltages and currents at ports in the cavity, when the ports are treated as electrically small antennas.

In this paper, we formulate and investigate the RCM as it applies to cases in which the ports are apertures in cavity walls. 
The aperture is assumed to be illuminated on one side by a plane electromagnetic wave.
We then distinguish between the radiation problem, where the aperture radiates into free space, and the cavity 
problem, where the aperture radiates into a closed electromagnetic (EM) environment. 
The solution of the problem in the cavity case is then given in terms of the free-space solution 
and a fluctuation matrix based on random matrix theory (RMT).
Thus, there is a clear separation between the system specific aspects of the aperture, in terms 
of the radiation admittance, and the cavity in terms of the fluctuation matrix. 

We illustrate our method by focusing on the problem of an electrically narrow aperture, for which the radiation admittance can be easily calculated 
numerically. 
We then generalize our result to the interesting case where a resonant mode of the aperture is excited. 
In this case the statistical properties of the aperture-cavity system can be given in a general universal form.

Our results build on previous work on apertures.
In particular, rectangular apertures have a very long research tradition in EM theory \cite{Harrington_1982}, and 
continue to be a topic of interest \cite{Lomakin2007}. 
The first self-consistent treatments have been carried out 
by Bethe \cite{Bethe_1944}, Bouwkamp \cite{1139646}, and, later, by Schwinger in aperture scattering \cite{Levine_1948_1}, and Roberts \cite{Roberts_1987}. 
Subsequent work on apertures is due to Ishimaru \cite{Ishimaru_1962}, Cockrell \cite{Cockrell_1976}, 
Harrington \cite{Harrington, Harrington_1982, Harrington_1985, Wang_1990}, and Ramat-Sahmii \cite{Rahmat_Samii_1977, Butler_1978}, 
among other investigators. 

Our results are of interest for the physical characterization of the radiation coupled into complex cavities such as reverberation chambers (RC) 
- which is known to be an extremely complicated problem even challenging classical EMC techniques \cite{Hill_Ma_1994,Ladbury_2002_symp,Fedeli_2009} 
- for understanding interference in metallic enclosures, as well as for modeling and predicting radiated emissions in complicated environments.

The paper is organized as follows. In Sec. II. we introduce the general model for the cavity backed aperture, and we describe the way the RCM models the 
cavity. In Sec. III. we apply the formulation of Sec. II. to large aspect ratio, rectangular apertures; evaluating the elements of the admittance matrix 
and computing the power entering a cavity with a rectangular aperture. In this section we also develop a simple formula for the power entering a low loss 
cavity with isolated resonances. Sec. IV. describes an extension of the model that accounts for the coupling of power through an aperture, into a cavity, and 
to an antenna in the cavity. Simple formulas for the high-loss case, low-loss, isolated resonance are developed.
\section{Random coupling model for apertures}\label{sec:RCM}
\begin{figure}[t]
\psfrag{r}{$\rho$}
\psfrag{x}{$\textbf{x}$}
\psfrag{xp}{$\textbf{x}_{\perp}$}
\psfrag{vc}{$V^{cav}$}
\psfrag{dvc}{$\partial V^{cav}$}
\psfrag{dva}{$\partial V^{ap}$}
\psfrag{p}{$P$}
\psfrag{zl}{$Z_L$}
\centering
\includegraphics[width=0.37 \textwidth]{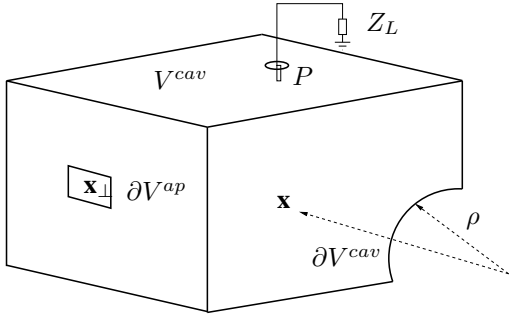}
\caption{\label{fig:cav_geo} Geometry of a complex 3D cavity with volume $V^{cav}$, boundary $\partial V^{cav}$, exhibiting wave-chaotic ray trajectories 
in the semiclassical limit: energy enters the cavity through an electrically large aperture $\partial V^{ap}$. Also shown is a point-like port to be considered in 
section \ref{sec:cav_port_ap}.}
\end{figure}

The random coupling model was originally formulated to model the impedance matrix of quasi-two-planar cavities with single \cite{zao1}, and multiple \cite{zao2} point-like ports \cite{Gradoni2014606}. 
In this section, we develop the model for three-dimensional \emph{irregular} enclosures excited through apertures 
\cite{Wiener_1953}.  
As depicted in Fig. \ref{fig:cav_geo}, we consider a cavity with a planar aperture in its wall through which 
our complex electromagnetic system is accessed from the outside. 
The size and shape of the aperture are important in our model, their specification constitutes ``system specific'' information 
that is needed to implement the model.
The cavity that we consider in our studies is an electrically large enclosure with an irregular geometry. 
The irregularity is assumed to be such that ray trajectories within the cavity are chaotic throughout \cite{Stock_1999}.
Generally speaking, typical cavities have this feature; particularly those with curved walls and/or with contents that scatter radiation in multiple directions. 
The consequences of assuming trajectories are chaotic is that the spectra of modes of the cavity have universal statistical properties that are modeled 
by random matrix theory (RMT) \cite{Mehta_2004}.
Experiments have been carried out to test the predictions of RCM for quasi-two-dimensional complex enclosures coupled through electrically small (point-like) 
ports \cite{Hemmady_2005,Zheng_PRE_2006,Hemmady_2006}. 
Experiments have also been carried out in three-dimensional enclosures, and with electrically-large ports \cite{Hemmady_2012,Drikas_2014}.

When apertures are considered, the port approximation invoked in the original derivation of the RCM is no longer valid, hence we need to consider the  
field distribution on the aperture surface. We consider a planar aperture, i.e., one that is not subject to boundary curvature.
This is consistent with real-world situations such as 3D reverberation chambers (RCs) \cite{Nested01TEMC_2003}, 
where the aperture is generally in a planar boundary. 

We first treat the aperture as if it existed in a metal plate separating two infinite half spaces. We refer to this as the ``free-space'' situation or ``radiation'' case. 
Suppose the port is treated as an aperture in a planar conductor whose surface normal $\hat{n}$ is parallel to the z-axis. 
The components of the fields transverse to $z$ in the aperture can be expressed as a superposition of modes (an example is the set of modes of a waveguide with the same cross sectional shape 
as the aperture)
\begin{equation}\label{eqn:Et_rad_aperture}
 \textbf{E}_t = \sum_s V_s \textbf{e}_s \left ( \textbf{x}_{\perp} \right ) \,\, ,
\end{equation} 
and
\begin{equation}\label{eqn:Ht_rad_aperture}
 \textbf{H}_t = \sum_s I_s \hat{n} \times \textbf{e}_s \left ( \textbf{x}_{\perp} \right ) \,\, ,
\end{equation}
where $\textbf{e}_s$ is the basis mode, (having only transverse fields) normalized such that $\int \, d^2 \textbf{x}_{\perp} \left | \textbf{e}_s \right |^2 = N_s$ and $\hat{n}$ 
is the normal, which we take to be in the $z$-direction.
In the radiation case we solve Maxwell`s equations in the half space $z > 0$ subject to the boundary conditions that 
$\textbf{E}_t  =  0$ on the conducting plane except at the aperture where it is given by (\ref{eqn:Et_rad_aperture}).  
We do this by removing the conducting plane, adding a magnetic surface current density $2 \delta \left ( z \right ) \hat{n} \times \textbf{E}_t$ to Faraday`s law, and solving Maxwell`s equations in the 
whole space $- \infty < z < \infty$ in the Fourier domain. 
For this problem, the transverse components of the electric field are odd functions of $z$ with a jump equal to twice (\ref{eqn:Et_rad_aperture}) at the location of the aperture; 
thus satisfying the boundary condition for the half space problem. 
We then evaluate the transverse components of the magnetic field on the plane $z = 0$, and project them on to the basis 
$\hat{n} \times \textbf{e}_s \left ( \textbf{x}_{\perp} \right )$ at the aperture to find the magnetic field amplitudes $I_s$ in (\ref{eqn:Ht_rad_aperture}). 
The result is a matrix relation between the magnetic field amplitudes and the electric field amplitudes in the aperture,
\begin{equation}\label{eqn:Y_aperture}
 I_s = \sum_{s^{'}} Y^{rad}_{s s^{'}} \left ( k_0 \right ) V_{s^{'}} \,\, ,
\end{equation}
where $k_0 = \omega / c$, $\omega$ is the frequency of excitation and we have adopted the phasor convention $\exp \left ( - i \omega t \right )$. 
Here, the radiation admittance matrix $Y^{rad}_{s s^{'}} $ is determined from the Fourier transform solution for the fields, and is given in terms of a 
three dimensional integral over wave numbers,
\begin{equation}\label{eqn:Yrad_aperture}
 Y^{rad}_{s s^{'}} \left ( k_0  \right ) = \sqrt{\frac{\mu}{\epsilon}} \int \frac{d^3 k}{\left ( 2 \pi \right )^3} 
 \frac{2 i k_0}{k^2_0 - k^2} \,\, \tilde{\textbf{e}}^{*}_s \cdot \ushortdw{\Delta}_{Y^{rad}} \cdot \tilde{\textbf{e}}_{s^{'}} \,\, ,
\end{equation}
where the dyadic tensor
\begin{equation}\label{eqn:Yrad_kern}
\begin{split}
 \ushortdw{\Delta}_{Y^{rad}} =  \frac{\textbf{k}_{\perp} \textbf{k}_{\perp}}{k_{\perp}^2} + 
 \left ( \frac{k^2 - k^2_{\perp}}{k^2} + \frac{\left ( k^2_0 - k^2 \right ) k^2_{\perp}}{k^2 k^2_0} \right ) 
 \\ \frac{\left ( \textbf{k} \times \hat{n} \right ) \left ( \textbf{k} \times \hat{n} \right )}{k_{\perp}^2} \,\, ,
 \end{split}
\end{equation}
is responsible for coupling two arbitrary modes of the aperture, and 
\begin{equation}\label{eqn:es_Fourier}
 \tilde{\textbf{e}}_{s} = \int_{aperture} \, d^2 \textbf{x}_{\perp} \, \exp \left ( - i \textbf{k}_{\perp} \cdot \textbf{x}_{\perp} \right ) \textbf{e}_{s} \,\,,
\end{equation}
is the Fourier transform of the aperture mode. The elements of the radiation admittance are complex quantities. 
The residue at the pole $k = k_0$ in (\ref{eqn:Yrad_aperture}) gives the radiation conductance 
\begin{equation}\label{eqn:Grad_aperture}
 G^{rad}_{s s^{'}} \left ( k_0  \right ) = \sqrt{\frac{\epsilon}{\mu}} \int \frac{k^2_0 d \Omega_k}{8 \pi^2} \,\, 
 \tilde{\textbf{e}}^{*}_s \cdot \ushortdw{\Delta}_{G^{rad}} \cdot \tilde{\textbf{e}}_{s^{'}} \,\, , 
\end{equation}
where $\Omega_k$ is the two dimensional solid angle of the wave vector $\textbf{k}$ to be integrated over $4 \pi$, and where there appears a modified 
dyadic tensor 
\begin{equation}\label{eqn:DeltaGrad}
 \ushortdw{\Delta}_{G^{rad}} = \left [ \left ( \textbf{k}_{\perp} \textbf{k}_{\perp} \right ) / k_{\perp}^2 \right ] + 
 \left [ \left ( \textbf{k} \times \hat{n} \right ) \left ( \textbf{k} \times \hat{n} \right ) / k_{\perp}^2 \right ] \,\, .
\end{equation}
The radiation conductance is frequency dependent through $k_0$. We note that there is an implicit $k_0$ dependence through the Fourier transforms of the 
aperture modes, and we set $\left | \textbf{k} \right | = k_0$ in (\ref{eqn:es_Fourier}) and (\ref{eqn:DeltaGrad}). 
The remaining part of (\ref{eqn:Yrad_aperture}), gives the radiation susceptance. 
Part of this can be expressed as a principle part integral of the radiation conductance. 
However, there is an additional inductive contribution ($Y \propto k_0^{-1}$) (which we term the magnetostatic conductance) coming from the last term in the 
parentheses in (\ref{eqn:Yrad_kern}) that contains a factor that cancels the resonant denominator in (\ref{eqn:Yrad_aperture}).
The reactive response of the aperture can be expressed in terms of the Cauchy principal value of the radiation conductance 
(\ref{eqn:Yrad_aperture}), and the previously mentioned inductive contribution, yielding
\begin{equation}\label{eqn:Brad_aperture}
 B^{rad}_{s s^{'}} \left ( k_0  \right ) =  P \int^{\infty}_{0} \frac{2 k_0 d k}{\pi \left ( k^2_0 - k^2 \right )} \,\, G^{rad}_{s s^{'}} \left ( k \right ) + B^{ms}_{s s^{'}} \,\, ,
\end{equation}
where $B^{ms}_{s s^{'}}$ stands for magnetostatic conductance, defined as 
\begin{equation}\label{eqn:Bms_aperture}
 B^{ms}_{s s^{'}} \left ( k_0  \right ) = \frac{2}{k_0} \sqrt{\frac{\epsilon}{\mu}} \int \frac{d k^3}{\left ( 2 \pi \right )^3}  \,\, 
 \tilde{\textbf{e}}^{*}_s \cdot \ushortdw{\Delta}_{B^{ms}} \cdot \tilde{\textbf{e}}_{s^{'}} \,\, , 
\end{equation}
and where $\ushortdw{\Delta}_{B^{ms}} =  \left [\left ( \textbf{k} \times \hat{n} \right ) \left ( \textbf{k} \times \hat{n} \right ) \right ] / k^2$.

We now repeat the process, but assume that the aperture opens into a cavity rather than an infinite half space, the radiation admittance (\ref{eqn:Yrad_aperture}) will be replaced by a cavity admittance.  
In appendix it is shown that under the assumptions that the eigenmodes of the closed cavity can be replaced by superpositions of random plane waves
(Berry's hypothesis \cite{h01TEMC_1998, Gradoni2009IEEE}), 
and the spectrum of the cavity eigenmodes can be replaced by one corresponding to a random matrix from the gaussian orthogonal ensemble (GOE) \cite{Mehta_2004}, 
the statistical properties of the cavity admittance can be represented as 
\begin{equation}\label{eqn:Ycav_fluct}
\ushortdw{Y}^{cav} = i \ushortdw{B}^{rad} + \left [ \ushortdw{G}^{rad} \right ]^{1/2} \cdot 
 \ushortdw{\xi} \cdot \left [ \ushortdw{G}^{rad} \right ]^{1/2} \,\, ,
\end{equation}
and the matrix $\ushortdw{\xi}$ is a universal fluctuation matrix defined as
\begin{equation}\label{eqn:universal_fluct}
     \ushortdw{\xi} = \frac{i}{\pi} \sum_n \frac{ \underline{\Phi}_n \underline{\Phi}^{T}_n}{\mathcal{K}_0 - \mathcal{K}_n + i \alpha} \,\, ,
\end{equation}
where $\mathcal{K}_n$ is a set of eigenvalues of a random matrix drawn from the GOE \cite{Mehta_2004}. 
These matrices have positive and negative eigenvalues that fall in a symmetric range about zero. 
The mean spacing of the eigenvalues near zero can be adjusted by scaling the size of the matrix elements. Here we assume 
that this has been done such that the mean spacing of the $\mathcal{K}_n$ near zero is unity. The quantity $\mathcal{K}_0$ 
represents the deviation of the excitation frequency $\omega_0$ from some reference value $\omega_{ref}$ placed in the band 
of interest
\begin{equation}\label{eqn:k_0_deviation}
 \mathcal{K}_0 = \frac{\omega_0 - \omega_{ref}}{\Delta \omega} \,\, ,
\end{equation}
where $\Delta \omega$ is the mean spacing between the resonant frequencies of modes of the actual cavity in the range of $\omega_{ref}$.
The resonant frequencies of a particular realization are thus given by $\omega_n = \omega_{ref} + \Delta \omega \mathcal{K}_n$.
The vector $\underline{\Phi}_n$ is composed of independent, zero mean, unit variance Gaussian random variables. The quantity $\alpha$ 
describes the average loss in the cavity and is related to the finesse parameter $\mathcal{F}$, widely used in optical cavities and photonic 
lattices \cite{born1999principles}
\begin{equation}
 \alpha = \mathcal{F}^{-1} = \frac{\omega_0}{2 Q \Delta \omega} \,\, .
\end{equation} 
Thus, according to the RCM, two quantities are required to specify the properties of the cavity: the $Q$ width $\omega_0 / Q$ and 
the mean spacing $\Delta \omega$ between nearest neighbor resonances (cavity modes).

The quantity $\alpha$ determines whether the cavity is in the high loss ($\alpha >1$) or low loss ($\alpha <1$) regime. More precisely, 
it establishes how much the individual resonances overlap. Basically, if $\alpha >1$ the $Q$-width of a resonance, $\omega_0 / Q$, exceeds the spacing between resonances $\Delta \omega$. 
A feature of chaotic cavities is that each mode in a given frequency band has essentially the same $Q$-width. This is a consequence of the ergodic 
nature of the ray trajectories that underlie the mode structure. Each mode has wave energy distributed throughout the cavity such that losses 
are approximately the same for each mode. 

\begin{figure}
          \psfrag{x}{$x$}
          \psfrag{y}{$y$}
          \psfrag{z}{$z$}
          \psfrag{L}{$L$}
          \psfrag{W}{$W$}
          \psfrag{psi}{$\phi_p$}
          \psfrag{tv}{$\hat{\theta}$}
          \psfrag{pv}{$\hat{\phi}$}
          \psfrag{t}{$\theta$}
          \psfrag{p}{$\phi$}
          \psfrag{hinc}{$\textbf{h}^{inc}$}
          \psfrag{k}{$\textbf{k}^{inc}$}
          \includegraphics[width=0.28 \textwidth]{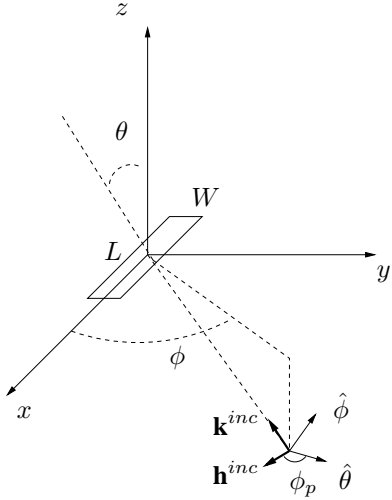}
          \centering
          \caption{\label{fig:ap_geo} Geometry of a 2D regular (narrow) aperture illuminated by an external plane-wave, radiating in free-space.}
\end{figure}

We now describe how relations (\ref{eqn:Y_aperture}) and (\ref{eqn:Ycav_fluct}) are used to determine the coupling of our system to external radiation.
Specifically, we consider the framework of Fig. \ref{fig:ap_geo}, where the aperture is illuminated by a plane wave incident with a wave vector $\textbf{k}^{inc}$  
and polarization of magnetic field $\textbf{h}^{inc}$ that is perpendicular to $\textbf{k}^{inc}$.  
We again consider the aperture to be an opening in a conducting plane at $z=0$, with radiation incident from $z=- \infty$.
We imagine writing the fields for $z < 0$ as the sum of the incident wave, the wave that would be specularly reflected from an infinite 
planar surface, and a set of outgoing waves associated with the presence of the aperture.  
The incident and specularly reflected waves combine to produce zero tangential electric field on the plane $z = 0$.  
Thus, the outgoing waves for $z < 0$ associated with the aperture can be expressed in terms of the electric fields in the aperture just as the outgoing waves for $z > 0$ can, 
the two cases being mirror images.  So, relation (\ref{eqn:Et_rad_aperture}) continues to represent the tangential electric fields in the plane $z = 0$.  
For the magnetic field we have separate expansions for $z > 0$ and $z < 0$.  
The electric field amplitudes are then determined by the condition that the magnetic fields are continuous in the aperture at $z = 0$.  For $z =0+$ we have 
$\textbf{H}^{>}_t = \sum_s I^{>}_s \hat{n} \times \textbf{e}_s \left ( \textbf{x}_{\perp} \right )$, with $\ushortw{I}^{>} = \ushortdw{Y}^{>} \cdot \ushortw{V}$, 
where $\ushortdw{Y}^{>}$ is either the radiation admittance matrix (\ref{eqn:Yrad_aperture}) or the cavity admittance matrix (\ref{eqn:Ycav_fluct}) depending on the circumstance. 
For $z = 0^{-}$ we have $\textbf{H}^{<}_t = \sum_s I^{<}_s \hat{n} \times \textbf{e}_s \left ( \textbf{x}_{\perp} \right ) + 
2 \textbf{h}^{inc} \exp \left [ i \textbf{k}^{inc} \cdot \textbf{x}_{\perp}\right ]$, where $\ushortw{I}^{<} = - \ushortdw{Y}^{rad} \cdot \ushortw{V}$ (the minus sign accounts for the 
mirror symmetry) and the factor of two multiplying the incident field comes from the addition of the incident and specularly reflected magnetic fields. 
Projecting the two magnetic field expressions on the aperture basis, and equating the amplitudes gives
\begin{equation}\label{eqn:radiation_aperture}
 \left ( \ushortdw{Y}^{>} + \ushortdw{Y}^{rad} \right ) \cdot \ushortw{V} = 2 \ushortw{I}^{inc} \,\, ,
\end{equation}
where 
\begin{equation}
  I_s^{inc} = - \hat{n} \cdot \tilde{\textbf{e}}_s \left ( - \textbf{k}^{inc}_{\perp} \right ) \times \textbf{h}^{inc} 
 \end{equation}
and $\tilde{\textbf{e}}_s$ is the Fourier transform of the aperture electric field and is defined in (\ref{eqn:es_Fourier}). 
Equation (\ref{eqn:radiation_aperture}) can be inverted to find the vector of voltages 
\begin{equation}\label{eqn:voltage}
 \ushortw{V} = \left ( \ushortdw{Y}^{rad} + \ushortdw{Y}^{>} \right )^{-1} \cdot 2 \ushortw{I}^{inc} \,\, ,
\end{equation} 
and then the net power passing through the aperture is given by
\begin{equation}\label{eqn:power_back}
 P_t = \frac{1}{2} \, \Re \left ( \ushortw{V}^{*} \cdot \ushortdw{Y}^{>} \cdot \ushortw{V} \right ) \,\, ,
\end{equation}
where $\ushortdw{Y}^{>}$ can be either a radiation (free-space) admittance (\ref{eqn:Yrad_aperture}) or a cavity admittance matrix (\ref{eqn:Ycav_fluct}).
\section{Rectangular apertures}
We consider rectangular apertures and select a basis for representation of the tangential fields in the aperture 
in (\ref{eqn:Et_rad_aperture}) and (\ref{eqn:Ht_rad_aperture}). One choice for the basis are the modes of a waveguide with a rectangular 
cross section. These can be written either as a sum of TE and TM modes, or simply as a Fourier representation of the individual 
Cartesian field components \cite{Bunting2005} \cite{Bunting2006}.

In EMC studies, narrow apertures are often considered, as they frequently occur in practical EM scenarios. 
The aperture of Fig. \ref{fig:ap_geo} is elongated and thin: it has only one electrically large dimension, and the field component $\textbf{e}_{\left ( n, 0, x \right )} \approx 0$.
Hence, in the particular case of a rectangular aperture with $L \thicksim  \lambda \gg W$ and $W \rightarrow 0$, the field will be dominated by $TE_{n0}$ modes 
\begin{equation}\label{eqn:es_narrow_approx}
 \textbf{e}_{s} \approx  \frac{\sin \left [ k_n \left ( x + L / 2 \right ) \right ]}{N}  \hat{y} \,\, , 
\end{equation}
for $\left | y \right | < W / 2$, where $N=\sqrt{2/LW}$. 
Once the aperture field basis is specified, the procedure for calculating the cavity admittance matrix (\ref{eqn:Yrad_aperture}) is given by the following steps \cite{Gradoni_EMCRoma_2012}. 
First, calculate the Fourier transform $\tilde{\textbf{e}}_{s}$ (\ref{eqn:es_Fourier}). 
Second, use $\tilde{e}_{s}$ to calculate the radiation conductance (\ref{eqn:Grad_aperture}). 
Third, use $\tilde{e}_{s}$ to calculate the magnetostatic susceptance (\ref{eqn:Bms_aperture}).
Fourth, use above quantities to form the radiation susceptance $G^{rad}_{s s^{'}} + i B^{rad}_{s s^{'}}$.
Finally, use the so-formed radiation admittance to generate the cavity admittance  (\ref{eqn:Ycav_fluct}), similar to \cite{zao2, s01PRE_2005}.
The conductance and susceptance have the property that off diagonal terms vanish if $n$ is odd and $n^{'}$ is even, and vice versa, 
due to the even and odd parity in $x$ of the basis modes. 
Further, in the limit of high frequency, $k_0 \gg k_n$, the components $G_{n n^{'}}$ have the limits 
\begin{equation}
 G^{rad}_{n n^{'}}  \rightarrow \,
  \left\{ \begin{array}{ll}
         \sqrt{\frac{\epsilon}{\mu}} & \,\,\,\, n = n^{'} \,\, , \\
         0 & \,\,\,\, n \neq n^{'} \,\, ,
        \end{array} \right .
\end{equation}
Thus, at high frequencies the admittance matrix $Y^{rad}_{s s^{'}}$ is diagonal and equal to the free space conductance as would 
be expected when the radiation wavelength is much smaller than the aperture size.

\begin{figure}[t]
\centering
\includegraphics[width=0.4 \textwidth]{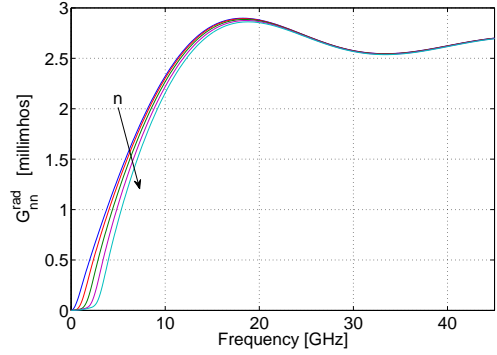}
\caption{\label{fig:Gnn_1_7} Diagonal radiation conductance for $n=1\ldots 5$, for a narrow aperture $L = 25$ cm $\times$ $W = 2$ cm. In the inset: 
low frequency (cutoff) detail of the diagonal elements with an off-diagonal element.}
\end{figure}

\begin{figure}[t]
\centering
\includegraphics[width=0.4 \textwidth]{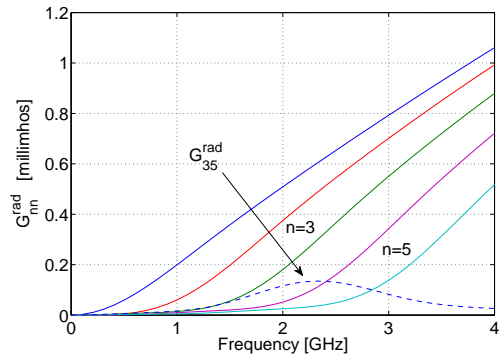}
\caption{\label{fig:Gnn_1_7_inset} Close-up of the diagonal radiation conductance in Fig. \ref{fig:Gnn_1_7}:
low frequency (cutoff) detail of the diagonal elements with an off-diagonal element (dashed line).}
\end{figure}

\begin{figure}[t]
\centering
\includegraphics[width=0.4 \textwidth]{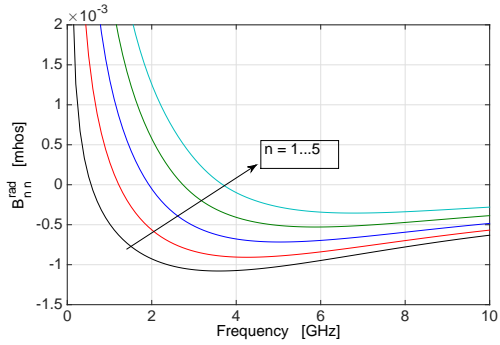}
\caption{\label{fig:Brad_1_5} Diagonal radiation susceptance for $n=1\ldots 5$, for a narrow aperture $L = 25$ cm $\times$ $W = 2$ cm.}
\end{figure}

We have evaluated the elements of the radiation conductance matrix as functions of frequency for a rectangular aperture with dimensions 
$L=25$ cm $\times W = 2$ cm. Plots of these appear in Figs. \ref{fig:Gnn_1_7} and \ref{fig:Gnn_1_7_inset}. 
For the diagonal elements the conductance increases nonmonotonically from zero, and asymptotes to its free-space value. 
For the off-diagonal components (not shown) the conductance first rises and then falls to zero with increasing frequency. 
Combining numerical evaluations of the two contributions in (\ref{eqn:Brad_aperture}) yields the total susceptance. 
This is plotted for several diagonal elements in Fig. \ref{fig:Brad_1_5}, which 
have the feature that for some elements the susceptance passes through zero at a particular frequency. 
At these frequencies the conductances (Fig. \ref{fig:Gnn_1_7_inset}) also tend to be small. This indicates the presence of resonant modes of the aperture. 
At the frequencies of these modes the aperture will allow more power to pass through the plane of the conductor than would expected based 
on the aperture area. 
Our next step is to evaluate the power passing through the aperture when it is illuminated by a plane wave. 
This involves solving (\ref{eqn:voltage}) for the vector of voltages $\ushortw{V}$, and inserting these in (\ref{eqn:power_back}) for the power transmitted through the aperture. 
This will be done a number of times; first with $\ushortdw{Y}^{>} = \ushortdw{Y}^{rad}$ to determine the power passing through the aperture in the radiation case; 
then again with $\ushortdw{Y}^{>} = \ushortdw{Y}^{cav}$ to determine the power through the aperture when it is backed by a cavity. 
In the cavity case a number of realizations of the cavity admittance matrix will be considered, modeling cavities with different distributions of resonant modes.

\begin{figure}[t]
\centering
\includegraphics[width=0.4 \textwidth]{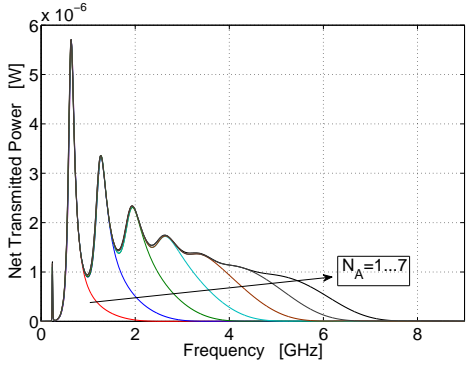}
\caption{\label{fig:power_rad_oblique} Frequency behavior of the net power transmitted by an aperture $\mathcal{A} = 0.25$ m $\times 0.02$ m in free space, 
$N_{\mathcal{A}} = 1, \ldots, 7$ modes, for an external plane wave of $h^{inc} = 1$ mA/m, at oblique incidence  $\phi = 0, \theta = \pi/4$, and polarization $\phi_p= \pi/2$.}
\end{figure}

Figure \ref{fig:power_rad_oblique} shows the frequency behavior of the transmitted power defined in (\ref{eqn:power_back}), with $\ushortdw{Y}^{>} = \ushortdw{Y}^{rad}$,  
for an oblique plane wave incident with angles $\phi_p=\pi/2$, $\phi =0$, and $\theta=\pi/4$, according to the coordinate frame of Fig. \ref{fig:ap_geo}. 
In Fig. \ref{fig:power_rad_oblique}, the net power is parameterized by the number of aperture modes included in the calculation, ranging from $N_{\mathcal{A}}=1$ to $N_{\mathcal{A}}=7$. 
As expected, the higher the frequency the greater the number of modes required to achieve an accurate prediction of the transmitted power. 
Interestingly, we notice the presence of a sharp peak at $600$ MHz, which is the slit resonance frequency, i.e., $f_A=c / (2 L)$, and other 
broader resonances located at $n \, f_A$.
At normal incidence, the peak of the resonance next to the sharp peak is reduced of a factor of $10$. This is confirmed in previous studies 
based on the transmission line model of a narrow aperture \cite{Warne_Chen_1990}.
Having computed the radiation conductance and susceptance for a narrow slit of dimensions $25$ cm $\times$ $2$ cm, we can investigate the effect of 
a wave-chaotic cavity backing the aperture by replacing $\ushortdw{Y}^{rad}$ by $\ushortdw{Y}^{cav}$ and exploiting the statistical model, (\ref{eqn:Ycav_fluct}). 
We first calculate distributions of the cavity admittance elements from the RCM by using a Monte Carlo technique and the radiation admittance of a rectangular aperture.
Numerical calculations of (\ref{eqn:Yrad_aperture}) and Monte Carlo simulation of (\ref{eqn:universal_fluct}) allow for generating an ensemble of cavity 
admittances of the form (\ref{eqn:Ycav_fluct}). In particular, we use the statistical method described in Refs. \cite{zao1}, and \cite{zao2}, with $N=7$ aperture modes, 
$M=600$ cavity modes, and a loss factor of $\alpha = 6$, simulating a chaotic cavity with high losses, to create the bare 
fluctuation matrix (\ref{eqn:universal_fluct}). 
Here, we repeat the simulation of (\ref{eqn:universal_fluct}) $800$ times to create an ensemble of fluctuation matrices.
By virtue of its construction, the average cavity admittance equals the radiation matrix $\left < \ushortdw{Y}^{cav} \right > = \ushortdw{Y}^{rad}$. Further, in the 
high loss limit ($\alpha \gg 1$) fluctuations in the cavity matrix become small and $\ushortdw{Y}^{cav} \rightarrow \ushortdw{Y}^{rad}$. 
For finite losses the character of the fluctuations in the elements of the cavity admittance matrix change from Lorenzian at low loss ($\alpha \ll 1$) to Gaussian at 
high loss ($\alpha \gg 1$). 

\begin{figure}[t]
\centering
\includegraphics[width=0.4 \textwidth]{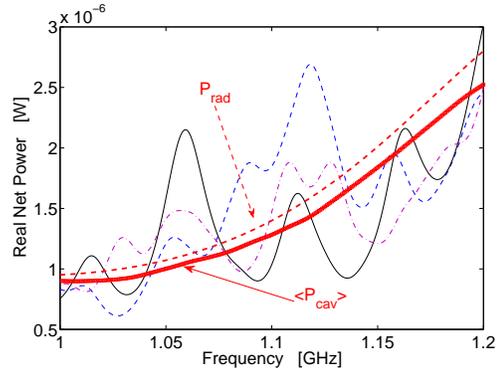}
\caption{\label{fig:power_cavity_a1} Comparison between the radiation (red dashed line), and the average (red squares) net power transmitted by an aperture of dimensions $\mathcal{A} = 0.25$ m $\times 0.02$ m, 
in the frequency range from $1$ GHz to $1.2$ GHz for an external incidence of $h^{inc} = 1$ A/m, direction of incidence  $\phi = 0, \theta = \pi/4$, polarization $\phi_p = \pi/2$. 
Thin solid (black), dashed (blue), and dash-dotted (purple) lines 
are reported to show $3$ independent cavity realizations as generated through a Monte Carlo method.
The thick solid (red) line indicates the ensemble average of the port power over $800$ cavity realizations. 
The thick dashed (red) line indicates the power radiated from the aperture in free-space.
Each cavity response is given by the superposition of 600 ergodic eigenmodes.
The chaotic cavity is modeled with $\alpha = 1.0$.}
\end{figure}

\begin{figure}[t]
\centering
\includegraphics[width=0.4 \textwidth]{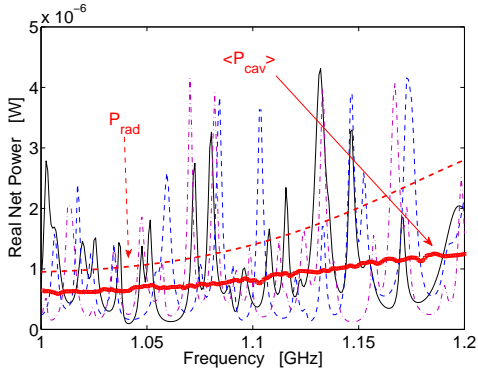}
\caption{\label{fig:power_cavity_a0p1} Comparison between the radiation (red dashed line), and the average (red squares) net power transmitted by an aperture of dimensions $\mathcal{A} = 0.25$ m $\times 0.02$ m, 
in the frequency range from $1$ GHz to $1.2$ GHz for an external incidence of $h^{inc} = 1$ A/m, direction of incidence  $\phi = 0, \theta = \pi/4$, polarization $\phi_p = \pi/2$. 
Thin solid (black), dashed (blue), and dash-dotted (purple) lines 
are reported to show $3$ independent cavity realizations as generated through a Monte Carlo method.
The thick solid (red) line indicates the ensemble average of the port power over $800$ cavity realizations. 
The thick dashed (red) line indicates the power radiated from the aperture in free-space.
Each cavity response is given by the superposition of 600 ergodic eigenmodes.
The chaotic cavity is modeled with $\alpha = 0.1$.}
\end{figure}

We now consider the net power coupled through an aperture that is backed by a wave-chaotic cavity. This involves evaluating (\ref{eqn:voltage}) and (\ref{eqn:power_back}) 
with $\ushortdw{Y}^{>} = \ushortdw{Y}^{cav}$ and with $\ushortdw{Y}^{cav}$ evaluated according to (\ref{eqn:Ycav_fluct}).
When this is done the frequency dependence of the net power acquires structure that is dependent on the density of modes in the cavity. 
This is illustrated in Figs. \ref{fig:power_cavity_a1} and \ref{fig:power_cavity_a0p1} where net power through the $0.25$ m by $0.02$ m rectangular aperture is plotted 
versus frequency in the range $1.0 < f < 1.2$ GHz for a few realizations of the fluctuating admittance matrix. 
In both cases the reference frequency in (\ref{eqn:k_0_deviation}) is set at $f_{ref} = \omega_{ref} / (2 \pi) = 1.1$ GHz, and the mean spacing between modes is set to be $\Delta f = \Delta \omega / (2 \pi) = 10$ MHz. 
Figure \ref{fig:power_cavity_a1} corresponds to a moderate loss case ($\alpha = 1.0$), and Fig. \ref{fig:power_cavity_a0p1} to a low loss case ($\alpha = 0.1$). 
Clearly, in the low loss case the peaks in power associated with different resonances are more distinct and extend to higher power. 
Also plotted in Figs. \ref{fig:power_cavity_a1} and \ref{fig:power_cavity_a0p1} are the average over $800$ realizations of the power coupled into the cavity and the power coupled through 
the aperture in the radiation case. Notice that both of these curves are smooth functions of frequency and that in the moderate loss case of Fig. \ref{fig:power_cavity_a1} the power averaged 
over many realizations $\left < P_{cav} \right >$ is only slightly less than transmitted power in the radiation case. 
In the low loss case of Fig. \ref{fig:power_cavity_a0p1} the average power coupled into the cavity is about half the value of the radiation case. 

The features of the coupled power in Fig. \ref{fig:power_cavity_a1} and \ref{fig:power_cavity_a0p1} 
can be captured by a simple model. 
We consider the frequency range in Fig. \ref{fig:power_rad_oblique} where the resonances in the aperture are isolated,
 roughly speaking $f < 3$ GHz. In this case, in a given narrow frequency range the aperture fields are dominated 
 by a single mode, 
and we can replace the matrix equations (\ref{eqn:radiation_aperture}-\ref{eqn:power_back}) with their scalar version
\begin{equation}\label{eqn:Pt_single_mode}
 P_t = 2 \left | I^{inc} \right |^2 \, \Re \left \{ \frac{Y^{>}}{\left | Y^{rad} + Y^{>} \right |^2} \right \} \,\, ,
\end{equation}
where $Y^{>}=Y^{rad}$ or $Y^{cav}$ depending on whether a cavity is present, 
$Y^{rad}=i B^{rad}+G^{rad}$ and $Y^{cav}=i B^{rad}+G^{rad} \xi$ with 
\begin{equation}\label{eqn:xi_single_mode}
 \xi = \frac{i}{\pi} \, \sum_{n^{'}} \, \frac{w^{2}_{n^{'}}}{\mathcal{K}_0 - \mathcal{K}_{n^{'}} + i \alpha} \,\, .
\end{equation}
Here $\mathcal{K}_0$, $\mathcal{K}_{n^{'}}$, and $\alpha$ are defined as before following (\ref{eqn:universal_fluct}) 
and the $w_{n^{'}}$ are a set of independent and identically distributed Gaussian random variables with 
zero mean and unit variance. The quantities $B^{rad}$ and $G^{rad}$ are properties of the aperture, and will be discussed below. 

The power transmitted through the aperture in the absence of a cavity is given by 
\begin{equation}
 P_t = \frac{\left | I^{inc} \right |^2}{2} \, \frac{G^{rad}}{\left | G^{rad} + i B^{rad} \right |^2}  \,\, .
\end{equation}
The aperture resonance occurs at a frequency $\omega_A$ where $B^{rad} \left ( \omega_A \right )=0$, 
corresponding to a peak in Fig. \ref{fig:power_rad_oblique}, and the coupled power at the peak is 
$P_A = P_t \left ( \omega_A \right ) = \left | I^{inc} \right |^2 / ( 2G^{rad})$, with $G^{rad} = G^{rad} (\omega_A)$. 
We can then express the frequency dependence of the power coupled through the aperture for 
frequencies near $\omega_A$ in the form of a Lorenzian resonance function 
\begin{equation}\label{eqn:Pt_rad_cav}
 P_t \left ( \omega \right ) = \frac{P_A}{1 + \Delta_A^2} \,\, ,
\end{equation}
where
\begin{equation}\label{eqn:Delta_aperture}
 \Delta_A = \frac{2 Q_A \left ( \omega - \omega_A \right )}{\omega_A} \,\, ,
\end{equation}
and the effective quality factor for the aperture is given by 
\begin{equation}
 Q_A = \frac{\omega_A}{2 G^{rad} \left ( \omega_A \right )} \left . \frac{d B^{rad}}{d \omega} \right |_{\omega_A} \,\, .
\end{equation}
When a cavity backs the aperture, $Y^{>}=Y^{cav}$ becomes a random frequency dependent function 
through the variable $\xi \left ( \omega \right )$. This quantity is frequency dependent through the 
denominators in (\ref{eqn:xi_single_mode}) and is random due to the random vectors of coupling coefficients $w_n$ 
and eigenvalues $\mathcal{K}_n$. The frequency scale for variation of $\xi \left ( \omega \right )$ is 
determined by the frequency spacing of modes of the cavity. In the typical case the frequency spacing of cavity modes  
is much smaller than that of aperture modes, as depicted in 
Figs. \ref{fig:power_cavity_a1} and \ref{fig:power_cavity_a0p1}. 
The behavior of the coupled 
power as a function of frequency will thus follow the envelope of the radiation case, with fluctuations on the frequency 
scale of the separation between cavity modes. 

This behavior can be captured in the simple mode if we assume the frequency is close to one of the poles ($n^{'}=n$) 
of (\ref{eqn:xi_single_mode}). 
Specifically we write 
\begin{equation}\label{eqn:xi_complex}
 \xi = i b + \frac{i \, w^2_n}{\pi \left ( \mathcal{K}_0 - \mathcal{K}_{n} + i \alpha \right )} \,\, ,
\end{equation}
where the term $i b$ is in the form of (\ref{eqn:xi_single_mode}) with the $n^{'}=n$ term removed and 
$\mathcal{K}_0 = \mathcal{K}_{n}$. Since $\alpha$ is assumed to be small it can be neglected in $b$ 
(making $b$ purely real) whereas it is retained in the second term in (\ref{eqn:xi_complex}) since we consider 
frequencies such that $\mathcal{K}_{0}-\mathcal{K}_{n}$ is small and comparable to $\alpha$. 
The statistical properties of the sum given by $b$ were detailed by Hart \emph{et al.} \cite{h01PRE_2009}. 
If we express $b$ in the form $b = \tan \psi$ then the PDF of $\psi$ is $\cos \psi / 2$ \cite{h01PRE_2009}. 
For now, since we are focusing on the behavior of individual realizations, we leave the value of $b$ unspecified. 

Having assumed the cavity response is dominated by a single resonance we can manipulate (\ref{eqn:Pt_single_mode}) 
into a general form 
\begin{equation}\label{eqn:Pt_alpha_alpha_n}
 P_t \left ( \omega \right ) = 
 P_A \frac{4 \alpha \alpha_A}{\left | \mathcal{K}_0 - \mathcal{K}^{'}_{n} + i \left ( \alpha + \alpha_{A} \right ) \right |^2} \,\, ,
\end{equation}
where $\mathcal{K}^{'}_{n} = \mathcal{K}_{n} - 2 \Delta^{'} \alpha_A$, $\Delta^{'} = \Delta_A + b / 2$, and 
\begin{equation}\label{eqn:alpha_A}
 \alpha_A = \frac{w_n^2}{\pi \left | 1 + 2 i \Delta^{'} \right |^2} \,\, .
\end{equation}
Here, the variables have the following interpretations: $\alpha_A$ is the ``external'' loss factor describing the damping 
of the cavity mode due to the aperture. Note, it is added to the internal loss factor $\alpha$ in the denominator 
of (\ref{eqn:Pt_alpha_alpha_n}). It is a statistical quantity, mainly through the Gaussian random variables $w_n$. 
This is responsible for variation in the height of the peaks in Fig. \ref{fig:power_cavity_a0p1}.
The quantity $\Delta^{'} = \Delta_A + b / 2$ represents the modification of the aperture resonance function 
(\ref{eqn:Delta_aperture}) by the reactive fields of the nonresonant cavity modes ($b / 2$). 
Note that it affects the external damping factor 
$\alpha_A$, which is largest when $\Delta^{'}=0$, i.e., when $\omega$ is near the aperture resonant frequency. 
Finally, $\mathcal{K}_n^{'}$ determines the shifted cavity mode frequency. 
That is, using definition (\ref{eqn:k_0_deviation}) the resonant cavity mode frequency, 
$\mathcal{K}_0 = \mathcal{K}_n^{'}$, becomes 
\begin{equation}
 \omega = \omega_n - 2 \Delta \omega \Delta^{'} \alpha_A \,\, .
\end{equation}
Equation (\ref{eqn:Pt_alpha_alpha_n}) implies that the power coupled into the cavity at frequency 
$\omega$ is bounded above by the power that can be transmitted through the aperture at the 
aperture resonance $\omega_A$. 
These powers are equal $P_t \left ( \omega \right ) = P_A$ if the mode is resonant, 
$\mathcal{K}_0=\mathcal{K}^{'}_n$, and the cavity is critically coupled, $\alpha = \alpha_A$. 
Note, however, that the coupled power in the cavity case (\ref{eqn:Pt_alpha_alpha_n}) can 
exceed the radiation case (\ref{eqn:Pt_rad_cav}) at the same frequencies if the aperture is off resonance 
$\Delta^{'} \left ( \omega \right ) \neq 0$. This is evident at the peaks of the coupled power in 
Fig. \ref{fig:power_cavity_a0p1}. 
Basically, what is happening is the cavity susceptance, which alternates in sign as frequency varies on the scale 
of the cavity modes, cancels the aperture susceptance, thus making the aperture resonant for frequencies away from 
the natural resonance $\omega_A$. 

\begin{figure}
   \centering
    \includegraphics[width=0.4 \textwidth]{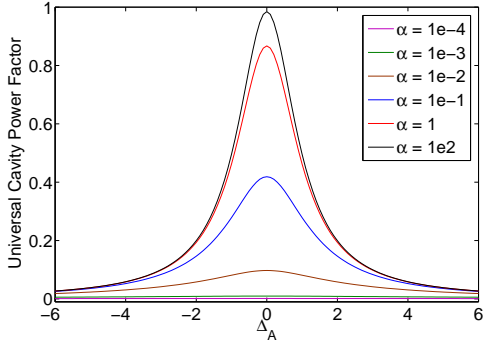}
    \caption{\label{fig:univ_factor_power} Universal aperture factor on $\Delta_A$, parameterized by $\alpha$.}
 \end{figure}

When the cavity loss parameter is small as in Fig. \ref{fig:power_cavity_a0p1} or (\ref{eqn:Pt_alpha_alpha_n}) 
there are large variations in coupled power as frequency is varied. A broad band signal would average 
over these variations. We can treat this case by computing the power coupled through the aperture averaged over realizations of the 
random variable $\xi$ defined in (\ref{eqn:xi_single_mode}). Such averages are shown in Fig. \ref{fig:power_cavity_a1} 
and \ref{fig:power_cavity_a0p1} based on Monte Carlo evaluations of the full system (\ref{eqn:power_back}). 
We perform this average in our simple model. 
A plot of a numerical evaluation of $\left < P_t \right > / P_A$ from (\ref{eqn:Pt_alpha_alpha_n}) as a function 
of $\Delta^{'}$ for different loss parameters $\alpha$ appears in Fig. \ref{fig:univ_factor_power}.
 Interestingly, a loss parameter that is as large as $\alpha = 1$ is sufficient to make the average power entering 
 a cavity $90 \%$ of that passing through an unbacked aperture.
\section{Cavity with both ports and apertures}\label{sec:cav_port_ap}
In the previous section we determined the properties of the power entering a cavity through an aperture. 
The contents of the cavity were treated as a distributed loss characterized by a single parameter $\alpha$. 
We will now extend this model to treat the case in which we identify a second port to the cavity that we will treat as an 
electrically small antenna. 
This port could be an actual port where a connection is made to the outside world as illustrated in Fig. \ref{fig:cav_geo}, 
or the port could represent the pin of a circuit element on which the voltage is of interest. 

We consider the configuration of Fig. \ref{fig:cav_geo}, where we have the joint presence of apertures and ports. 
We have previously considered the case of fields excited by a current distribution of the form \cite{Antonsen_2011}
\begin{equation}\label{eqn:J_multiport}
 \textbf{J} \left ( \textbf{x} \right ) = \sum_p \textbf{u}_p \left ( \textbf{x} \right ) I_p,\,\, ,
\end{equation}
where $\textbf{u}_p \left ( \textbf{x} \right )$ is a set of basis functions used to represent the current distribution 
in terms of a set of amplitudes $I_p$ which we call the port currents. The corresponding port voltages were defined in Ref. \cite{Antonsen_2011} as 
\begin{equation}\label{eqn:Vp_def}
 V_p = - \int d^3 x \, \textbf{u}_p \left ( \textbf{x} \right ) \cdot \textbf{E} \left ( \textbf{x} \right ) \,\, ,
\end{equation}
and as a result the power entering the cavity through the ports is $P_P = \Re \left \{ \sum_p V^{*}_p I_p \right \} / 2$. 

In analogy to our treatment of the aperture, we consider two cases: one in which the current distribution radiates into free-space and 
one in which the current distribution radiates into a cavity. The linear relationship between the port voltages $V_p$ and the port currents $I_p$ 
is then characterized by impedance matrices $Z^{rad}_{p p^{'}}$ and $Z^{cav}_{p p^{'}}$ depending on the case under consideration, where 
\begin{equation}\label{eqn:Vp_multiport}
 V_p = \sum_{p^{'}} Z^{rad/cav}_{p p^{'}} I_{p^{'}} \,\, .
\end{equation}

For the radiation case it was shown \cite{Antonsen_2011} 
\begin{equation}\label{eqn:Zrad_multiport}
 Z^{rad}_{pp^{'}} \left ( k_0 = \omega / c \right ) = \sqrt{\frac{\mu}{\epsilon}} \int \frac{d^3 k}{\left ( 2 \pi \right )^3} 
 \frac{i k_0}{k^2_0 - k^2} \tilde{\textbf{u}}_p \cdot \ushortdw{\Delta}_Z \cdot \tilde{\textbf{u}}^{*}_{p^{'}} \,\, .
\end{equation}
where $\tilde{\textbf{u}}_p \left ( k \right )$ is the Fourier transform of the basis function $\textbf{u}_p \left ( \textbf{x} \right )$, $k_0 = \omega / c$, and the dyadic 
$\ushortdw{\Delta}_Z$ is given by 
\begin{equation}
 \ushortdw{\Delta}_Z = \frac{\textbf{1} k^2 - \textbf{k} \textbf{k}}{k^2} + \frac{\textbf{k} \textbf{k}}{k^2 k^2_0} 
 \left ( k^2_0 - k^2 \right ) \,\, .
\end{equation}
The radiation impedance matrix can be decomposed $Z^{rad}_{pp^{'}} = R^{rad}_{pp^{'}} + i X^{rad}_{pp^{'}}$ where $R^{rad}_{pp^{'}}$ is the residue 
from the pole at $k = k_0$ in (\ref{eqn:Zrad_multiport}) (the radiation resistance)
\begin{equation}\label{eqn:Rrad_multiport}
 R^{rad}_{pp^{'}} \left ( k_0 \right ) = \Re \left ( Z^{rad}_{pp^{'}} \right ) = 
 \sqrt{\frac{\mu}{\epsilon}} \int \frac{k^2_0 d \Omega_k}{16 \pi^2} 
 \tilde{\textbf{u}}^{*}_p \cdot \frac{\textbf{1} k^2 - \textbf{k} \textbf{k}}{k^2} \cdot \tilde{\textbf{u}}^{*}_{p^{'}} \,\, ,
\end{equation}
and $X^{rad}_{pp^{'}}$ is the reactive contribution.  

For the cavity case it was shown 
\begin{equation}\label{eqn:Zcav_3D}
 \ushortdw{Z}^{cav} = i \Im \left \{ \ushortdw{Z}^{rad} \right \} + 
    \left [ \ushortdw{R}^{rad} \right ]^{1/2} \cdot \ushortdw{\xi} \cdot \left [ \ushortdw{R}^{rad} \right ]^{1/2} \,\, ,
\end{equation}
where the fluctuating matrix $\ushortdw{\xi}$ is defined in (\ref{eqn:universal_fluct}).

We are now in a position to describe a statistical model for a cavity including both an aperture and a localized current distribution. 
In this case we construct an input column vector $\ushortw{\phi}$ that consists of the aperture voltages and port currents and an output vector $\ushortw{\psi}$ 
that consists of the aperture currents and port voltages
 \begin{equation}
  \ushortw{\phi} = \left [ 
 \begin{array}{c}
    \ushortw{V}_{A}  \\
    \ushortw{I}_{P} 
 \end{array} \right ]
 \end{equation}
 and
 \begin{equation}
  \ushortw{\psi} = \left [ 
 \begin{array}{c}
    \ushortw{I}_{A}  \\
    \ushortw{V}_{P} 
 \end{array} \right ]
\end{equation} 
where $\ushortw{V}_{A,P}$ are the aperture and port voltages, and $\ushortw{I}_{A,P}$ are the aperture and port currents.
These are then related by a hybrid matrix $\ushortdw{T}$, $\ushortw{\psi} = \ushortdw{T} \cdot \ushortw{\phi}$, where
\begin{equation}\label{eqn:RCM_hybrid}
   \ushortdw{T} = i \, \textrm{Im} \left ( \ushortdw{U} \right ) + \left [ \ushortdw{V} \right ]^{1/2} \cdot 
   \ushortdw{\xi} \cdot \left [ \ushortdw{V} \right ]^{1/2} \,\, .
\end{equation}
Here the matrices $\ushortdw{U}$ and $\ushortdw{V}$ are block diagonal, viz., 
\begin{equation}
 \ushortdw{U} =  \left [ 
 \begin{array}{cc}
    \ushortdw{Y}^{rad} & 0  \\
    0 & \ushortdw{Z}^{rad} 
 \end{array} \right ] \,\, ,
\end{equation}
and $\ushortdw{V} =  \Re \left [ \ushortdw{U} \right ]$.
The dimension of $\ushortdw{U}$ and $\ushortdw{V}$ is $\left ( N_s + N_p \right ) \times \left ( N_s + N_p \right )$, where $N_p$ is the number of port currents and 
$N_s$ is the number of aperture voltages. 
Here we have assumed that the ports and apertures are sufficiently separated such that the off diagonal terms in $\ushortdw{U}$, describing the direct excitation 
of port voltages by aperture voltages, and aperture currents by port currents, are approximately zero. 
This assumption can be released in case of direct illumination between, or proximity of, aperture and port through the short-orbit correction of the RCM, 
which in this setting needs to be extended to cope with vector electromagnetic fields \cite{h01PRE_2009,JHY2010}.
In the simples case we can take the square root of $\ushortdw{V}$
\begin{equation}
 \left [ \ushortdw{V} \right ]^{1/2} =  \left [ 
 \begin{array}{cc}
    \left [ \ushortdw{G}^{rad} \right ]^{1/2} & 0  \\
    0 & \left [ \ushortdw{R}^{rad} \right ]^{1/2}
 \end{array} \right ] \,\, .
\end{equation}
At this point we specialize consideration to the case of a port that is an electrically small antenna characterized by a single current $I_p$ and single basis function 
in (\ref{eqn:J_multiport}). The matrices $\ushortdw{T}$, $\ushortdw{U}$, and $\ushortdw{V}$ then have dimension $(N_A + 1) \times (N_A + 1)$ and 
the matrix relations for the voltages and currents are more clearly expressed when separated into $N_A$ aperture equations and one  port equation. 
For the aperture equations we find that (\ref{eqn:radiation_aperture}) is replaced by 
\begin{equation}\label{eqn:aperture_equations}
 \left ( \ushortdw{Y}^{cav} + \ushortdw{Y}^{rad} \right ) \cdot \ushortw{V}_{A} + \left [  \ushortdw{G}^{rad} \right ]^{1/2} \cdot \ushortw{\xi}_{AP} \cdot \left ( R^{rad} \right )^{1/2} I_p = 2 \ushortw{I}^{inc} \,\, .
\end{equation}
For the port we assume the small antenna drives a load with impedance $Z_L$ such that $V_p = - Z_L I_p$. Then the port equation becomes 
\begin{equation}\label{eqn:port_equation}
 \left ( Z_L + Z^{cav} \right ) I_p + \left ( R^{rad} \right )^{1/2} \ushortw{\xi}^{T}_{AP} \cdot \left [ \ushortdw{G}^{rad} \right ]^{1/2} \cdot \ushortw{V}_{A} = 0 \,\, ,
\end{equation}
where $\ushortw{\xi}_{AP}$ is a column vector defined similarly to the $N_A \times N_A$ matrix $\ushortw{\xi}$ in (\ref{eqn:universal_fluct}). 
Finally, the scalar cavity impedance is defined in accord with (\ref{eqn:Zcav_3D})
\begin{equation}
 Z^{cav} = i X^{rad} + R^{rad} \xi \,\, ,
\end{equation}
with $\xi$ being a scalar version of the fluctuation matrix.
If we assume the aperture voltages are known, (\ref{eqn:port_equation}) can be solved for the current in the port, 
\begin{equation}\label{eqn:port_current_fluct}
 I_p = - \left ( Z_L + Z^{cav} \right )^{-1} \left ( R^{rad} \right )^{1/2} \ushortw{\xi}^{T}_{AP} \cdot \left [ \ushortw{G}^{rad} \right ]^{1/2} \cdot \ushortw{V}_{A} \,\, .
\end{equation}
Substituting the expression for the port current into the aperture equations (\ref{eqn:aperture_equations}) yields an equation of the form of (\ref{eqn:radiation_aperture}) 
for the aperture voltages 
\begin{equation}\label{eqn:port_radiation_aperture}
 \left ( \ushortdw{Y}^{rad} + \ushortdw{Y}^{cav '} \right ) \cdot \ushortw{V}_A = 2 \ushortw{I}^{inc} \,\, .
\end{equation}
Here the modified cavity admittance 
\begin{equation}\label{eqn:modified_Ycav}
 \ushortdw{Y}^{cav '} = \ushortdw{Y}^{cav} - \frac{R^{rad}}{\left ( Z_L + Z^{cav} \right )} \left [  \ushortdw{G}^{rad} \right ]^{1/2} \cdot \ushortw{\xi}_{AP} \cdot 
 \ushortw{\xi}^{T}_{AP} \cdot \left [ \ushortdw{G}^{rad} \right ]^{1/2} \,\, ,
\end{equation}
includes the effect of the current induced in the port antenna on the magnetic fields at the aperture. In the high loss limit ($\alpha > 1$)  or if the antenna load 
impedance is large this last term is small, $\ushortdw{Y}^{cav '} \approx \ushortdw{Y}^{cav}$, and the amount of power coupled through the aperture is unaffected by the presence 
of the antenna. Once (\ref{eqn:port_radiation_aperture}) is inverted to find the aperture voltages, $\ushortw{V}_A$, (\ref{eqn:modified_Ycav}) can be used to find 
the current induced in the antenna, and the relation $V_p = - Z_L I_p$ can be used to find the port voltage. The power dissipated in the load is then $P_L = R_L \left | I_p \right |^2 / 2$. 

We first consider the statistical properties of the fluctuating port voltage $V_p$ in the high loss limit. As mentioned, in this case $\ushortdw{Y}^{cav '} \approx \ushortdw{Y}^{cav}$, and 
further $\ushortdw{Y}^{cav} \approx \ushortdw{Y}^{rad}$. Thus, the aperture voltages $\ushortw{V}_A$ determined by (\ref{eqn:port_radiation_aperture}) will be the same as those 
determined in the radiation case of (\ref{eqn:voltage}) and the power coupled into the cavity will have the character displayed in Fig. \ref{fig:power_rad_oblique}. 
Statistical variations in the port current are then determined by the column vector $\ushortw{\xi}_{AP}$ in (\ref{eqn:port_current_fluct}). 
(In the high loss limit, variations in $Z^{cav}$ become small, and we can replace $Z^{cav}$ with $Z^{rad}$ for the port antenna.)

The statistical properties of the elements of $\ushortw{\xi}^{T}_{AP}$ are the same as those of the off-diagonal elements of the general matrix $\ushortdw{\xi}$ 
defined in (\ref{eqn:universal_fluct}). In the limit of high loss the elements are complex, with independent real and imaginary parts, each of which are zero mean 
independent Gaussian random variables \cite{zao1,zao2}. 
The common variance for the real and imaginary parts is $\left ( 2 \pi \alpha \right )^{-1}$. Since according to (\ref{eqn:port_current_fluct}) the port current $I_p$ is a linear 
superposition of Gaussian random variables, it too has independent real and imaginary parts that are zero mean Gaussian random variables. What remains then is to 
calculate its variance. Specifically, we find 
\begin{equation}\label{eqn:variance_Ip}
 \left < \left | I_p \right |^2 \right > = \frac{R^{rad}}{\left | Z_L + Z^{rad} \right |^2} \frac{1}{\pi \alpha} \ushortw{V}^{*}_A \cdot \ushortdw{G}^{rad} \cdot \ushortw{V}_A \,\, .
\end{equation}
Similar arguments determine the statistics of the port voltage. Since the port voltage and current are complex with independent Gaussian distributed real and imaginary 
parts, their magnitudes will be Rayleigh distributed with a mean determined by (\ref{eqn:variance_Ip}).
Finally, we calculate the expected power coupled to the load $\left < P_L \right > = R_L \left < \left | I_p \right |^2 \right > / 2$,
\begin{equation}\label{eqn:power_port_load}
 \left < P_L \right > = \frac{R_L R^{rad}}{\left | Z_L + Z^{rad} \right |^2} \frac{1}{2 \pi \alpha} P_t \,\, ,
\end{equation}
where $P_t$ is the power transmitted through the aperture in the radiation case. We note that the ratio of the power coupled to the load and the power entering 
the cavity satisfies a relation identical to what we found previously for the case of localized ports \cite{gradoni2012PRE}. 

\begin{figure}[t]
\centering
\includegraphics[width=0.4 \textwidth]{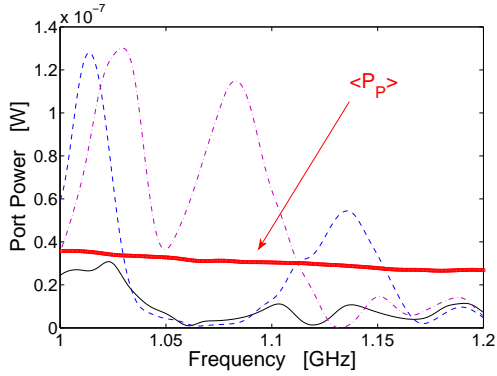}
\caption{\label{fig:P_P_alpha_1p0} Port power received by an antenna in the frequency range from $1$ GHz to $1.2$ GHz. 
The cavity is highly irregular and overmoded, with loss factor of $\alpha = 1.0$. The aperture is of dimensions $L = 0.25$ m $\times W = 0.02$ m, and the external plane-wave 
we assumed has amplitude $\left | \textbf{h}^{inc} \right | = 1$ mA/m, and direction of incidence $\theta = \pi / 4$, $\psi = 0$, and $\phi = 0$. 
Thin solid (black), dashed (blue), and dash-dotted (purple) lines indicate the port power for $3$ independent cavity realizations, 
while the thick solid (red) line indicates the ensemble average of the port power over $800$ cavity realizations.
Each cavity response is given by the superposition of 600 ergodic eigenmodes.}
\end{figure}

\begin{figure}[t]
\centering
\includegraphics[width=0.4 \textwidth]{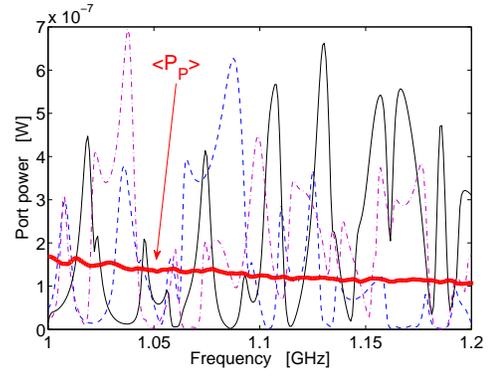}
\caption{\label{fig:P_P_alpha_0p1} Port power received by an antenna in the frequency range from $1$ GHz to $1.2$ GHz. 
The cavity is highly irregular and overmoded, with loss factor of $\alpha = 0.1$. The aperture is of dimensions $L = 0.25$ m $\times W = 0.02$ m, and the external plane-wave 
we assumed has amplitude $\left | \textbf{h}^{inc} \right | = 1$ mA/m, and direction of incidence $\theta = \pi / 4$, $\psi = 0$, and $\phi = 0$. 
Thin solid (black), dashed (blue), and dash-dotted (purple) lines 
indicate the port power for $3$ independent cavity realizations, while the thick solid (red) line indicates the ensemble average of the port power over $800$ cavity realizations.
Each cavity response is given by the superposition of 600 ergodic eigenmodes.} 
\end{figure}

The numerical computation of the \emph{exact} expressions (\ref{eqn:port_current_fluct}), (\ref{eqn:port_radiation_aperture}), and (\ref{eqn:modified_Ycav}) is now carried out. 
We now describe numerical solutions of the coupled port and aperture equations (\ref{eqn:aperture_equations}-\ref{eqn:port_radiation_aperture}). Elements of the 
matrix $\ushortdw{\xi}$, (\ref{eqn:universal_fluct}), vector $\ushortw{\xi}_{AP}$, and scalar $\xi$ were generated using the same Monte Carlo algorithm used to produce 
Figs. (\ref{fig:power_cavity_a1}) and (\ref{fig:power_cavity_a0p1}). 
Specifically, we generated $800$ realizations each with $600$ modes. The cavity was excited through a narrow aperture of dimensions 
$L=0.25$ m $\times W=0.02$ m, by a plane wave with amplitude $\left | \textbf{h}^{inc} \right | = 1$ mA/m, direction of incidence $\theta = 0$, and $\psi = \pi / 2$, with 
polarization $\phi = \pi / 2$, and with frequency swept from $1$ GHz to $1.2$ GHz. The receiving port was an antenna with free-space impedance 
$Z^{rad} = 30 - j 20$ $\Omega$ terminated by a $Z_L = 50$ $\Omega$ load. Figures \ref{fig:P_P_alpha_1p0} and \ref{fig:P_P_alpha_0p1} 
show the power coupled to the load for the case of loss factor 
$\alpha = 1.0$ and $\alpha = 0.1$ respectively. As in the cases of the power coupled into the cavity, Figs. \ref{fig:power_cavity_a1} and \ref{fig:power_cavity_a0p1}, 
the power to the load for individual realizations 
shows variations with frequency which are more pronounced in the low-loss case than in the moderate loss case. 
Figures \ref{fig:P_P_alpha_1p0} and \ref{fig:P_P_alpha_0p1} also display the power to the 
load averaged over the $800$ realizations. The amount of power reaching the load is generally in agreement with (\ref{eqn:power_port_load}). 
From Figs. \ref{fig:power_cavity_a1} and \ref{fig:power_cavity_a0p1} we see that the average power entering the cavity 
at $1.1$ GHz is $1.6$ $\times$ $10^{-6}$ W and $0.6$ $\times$ $10^{-6}$ W in the 
$\alpha = 1.0$ and $\alpha=0.1$ cases respectively. Equation (\ref{eqn:power_port_load}) then predicts for the power reaching the load $0.5$ $\times$ $10^{-7}$ W and 
$2.1$ $\times$ $10^{-7}$ W in these cases. Note that more power enters the cavity in the higher loss case, but more power reaches the load in the lower loss case.
In the low loss limit we can develop a formula analogous to (\ref{eqn:Pt_alpha_alpha_n}). We assume the elements of the $\ushortdw{\xi}$-matrix are dominated 
by a single cavity mode, and make the same approximations as the one leading to (\ref{eqn:Pt_alpha_alpha_n}). The results for the power coupled to the load is found to be 
\begin{equation}\label{eqn:PL_alpha_alpha_A}
 P_L \left ( \omega \right ) = 
 P_A \frac{4 \alpha_A \alpha_P}{\left | \mathcal{K}_0 - \mathcal{K}^{''}_{n} + i \left ( \alpha + \alpha_{A} + \alpha_P \right ) \right |^2} \,\, ,
\end{equation}
where $P_A$ is the power coupled through the aperture at resonance, $\alpha_A$ is the aperture loss factor defined in (\ref{eqn:alpha_A}), $\alpha_P$ is an 
analogously defined loss factor for the port 
\begin{equation}\label{eqn:alpha_P}
 \alpha_P = \frac{R_L R^{rad} w^2_P}{\pi \left | Z_L + i \left ( X^{rad} + R^{rad} b_P \right ) \right |^2} \,\, ,
\end{equation}
and $\mathcal{K}_0 = \mathcal{K}^{''}_{n}$ determines the shifted resonant frequency of mode-$n$, where 
\begin{equation}
 \mathcal{K}^{''}_{n} =  \mathcal{K}_{n} - 2 \delta^{'} \alpha_A - \alpha_P \left ( X^{rad} + X_L + R^{rad} b_P \right ) / R_L \,\, .
\end{equation}
In (\ref{eqn:PL_alpha_alpha_A}) the statistically fluctuating quantities $\alpha_P$ and $\alpha_A$ 
determine the peak power at resonance reaching the antenna. In the low loss case considered here, the peak power delivered to the load can be a substantial fraction 
of the power passing through the aperture at the aperture resonance, $P_A$.
\section{Conclusion}
We have developed a statistical model of the coupling of electromagnetic power through an aperture into a wave chaotic cavity. 
The formulation combines the deterministic, and system specific elements of the aperture and a receiving antenna in the cavity, 
with a statistical model for the modes of the cavity.
Particular attention has been focused on the case of a rectangular aperture, which has a set of well separated resonances. Power coupled through an 
aperture into free-space has peaks at frequencies corresponding to the modes of the aperture. When the aperture is backed by a cavity a new set 
of peaks appears at frequencies of modes of the cavity. These peaks can be as large as the free-space aperture resonance peaks. Simple formulas 
are derived that augment the detailed formulation, and that apply in the high-loss, low-loss, isolated resonance limit. 
These results are of interest for studying the coupling of an external radiation through apertures in complex cavities, such as RCs, and for practical 
scenarios of slotted enclosures populated by materials and electronic circuitry.
\section*{Acknowledgement}
This work was supported by the Air Force Office of Scientific Research FA95501010106 and the Office of Naval Research N00014130474.
\appendix \label{sec:appendix}
The admittance of a cavity excited through an aperture can be expressed by expanding the fields inside the cavity in a basis of electric and magnetic modes
\begin{equation}
 \textbf{E} = \sum_n \, V_n^{em} \textbf{e}_n^{em} \left ( \textbf{x} \right ) \,\, ,
\end{equation}
and
\begin{equation}
 \textbf{H} = \sum_n \, \left ( I_n^{em} \textbf{h}_n^{em} \left ( \textbf{x} \right ) + I_n^{ms} \textbf{h}_n^{ms} \left ( \textbf{x} \right ) \right ) \,\, .
\end{equation}
Here the electromagnetic modes satisfy the pair of equations, $-i k_n \textbf{e}_n^{em} = \nabla \times \textbf{h}_n^{em}$, and 
$i k_n \textbf{h}_n^{em} = \nabla \times \textbf{e}_n^{em}$ with the tangential components of the electric field equal to zero on the cavity boundary including the aperture. 
The magnetostatic modes are irrotational $\textbf{h}_n^{ms}=- \nabla \chi_n$, where the potential satisfies the Helmholtz equation $\left ( \nabla^2 + k_n^2 \right ) \chi_n =0$, 
with Neumann boundary conditions, $\left | \hat{n} \cdot \nabla \chi_n \right |_B =0$. 
The magnetostatic modes are needed to represent magnetic fields that have nonvanishing normal components at the aperture \cite{Bevensee1960}. 
It can be shown that all the magnetic field modes are orthogonal.

The mode amplitudes are determined by projecting Maxwell\'s equations onto the basis functions for each field type. 
The result of this action is an expression for the magnetic field amplitudes in the aperture that is equivalent to (\ref{eqn:Y_aperture}) 
except that the radiation admittance matrix is replaced by a cavity admittance matrix
\begin{equation}\label{eqn:Zcav_multiport} 
 Y^{cav}_{ss^{'}} \left ( k_0 \right ) = \sqrt{\frac{\epsilon}{\mu}} \sum_n \
 \left ( \frac{i k_0}{k^2_0 - k^2_n} \frac{w^{em}_{sn} w^{em}_{s^{'}n}}{V^{em}} + 
 \frac{i}{k_0} \frac{w^{ms}_{sn} w^{ms}_{s^{'}n}}{V^{ms}} \right ) \,\, ,
\end{equation}
where
\begin{equation}
 w^{\left ( \cdot \right )}_{sn} = \int_{aperture} d \textbf{x}_{\perp} \,  \textbf{e}_s \left ( \textbf{x}_{\perp} \right ) \cdot 
 \hat{z} \times \textbf{h}^{\left ( \cdot \right )}_n \,\, ,
\end{equation}
is the projection of the magnetic filed of the cavity mode onto the aperture field profile and
\begin{equation}
 V^{\left ( \cdot \right )} =  \int d \textbf{x} \, \left | \textbf{h}^{\left ( \cdot \right )}_n \right |^2 \,\, ,
\end{equation}
is a normalization factor for the eigenfunctions.  

Expression (\ref{eqn:Zcav_multiport}) is general and gives an exact expression for the admittance matrix of a lossless cavity in terms of the cavity modes. 
If we apply the random coupling hypothesis, we replace the exact eigenmodes with modes corresponding to random superpositions of plane waves. 
Specifically, near the plane $z = 0$ we write for the components of the electromagnetic eigenmodes transverse  to the z direction
\begin{equation}
 \textbf{h}^{em}_{n \perp} = \lim_{N \rightarrow \infty} \frac{2}{\sqrt{N}} \sum^N_{j=1} \, \textbf{b}_{j \perp} \cos \left ( \textbf{k}_j \cdot \hat{n} z \right ) 
 \cos \left ( \theta_j + \textbf{k}_j \cdot \textbf{x}_{\perp} \right ) \,\, ,
\end{equation}
where $\theta_j$ are uniformly distributed in the interval $\left [ 0, 2 \pi\right ]$, $\left | \textbf{k}_j \right | = k_n$, with the direction of $\textbf{k}_j$ uniformly  distributed over 
the half solid angle corresponding to $\textbf{k}_j \cdot \hat{n} > 0$, $\left | \textbf{b}_j \right | = 1$, with $\textbf{b}_j$ uniformly distributed in angle in 
the plane perpendicular to $\textbf{k}_j$. Except as mentioned, all random variables characterizing each plane wave are independent.  
A similar expression can be made for the scalar potential $\chi_n$ generating the magnetostatic modes. 
With eigenfunctions expressed as a superposition of random plane waves each factor $w_{sn^{(\cdot)}}$ appearing in (\ref{eqn:Zcav_multiport}) becomes 
a zero mean Guassian random variable. 
The correlation matrix between two such factors can then be evaluated by forming the product of two terms, averaging over the random variables 
parameterizing the eigenfunctions and taking the limit $N \rightarrow \infty$. 
We find for the electromagnetic modes the following expectation value
\begin{equation}\label{eqn:corr_em_mode_ampl} 
\begin{split}
 \left < \frac{w_{sn}^{(em)} w_{s^{'}n}^{(em)}}{V^{em}} \right > = \\  \int \, \frac{d \Omega_k}{4 \pi V} \, \tilde{\textbf{e}}^{*}_s \cdot 
 \left [ \frac{\textbf{k}_{\perp} \textbf{k}_{\perp}}{k_{\perp}^2} + 
 \left ( \frac{k^2_{\parallel}}{k^2} \right ) \frac{\left ( \textbf{k} \times \hat{n} \right ) \left ( \textbf{k} \times \hat{n} \right )}{k_{\perp}^2} \right ]  \cdot \tilde{\textbf{e}}_{s^{'}} \,\, ,
 \end{split}
\end{equation}
Here $\left | \textbf{k} \right | = k_n$, $\Omega_n$ represents the spherical solid angle of $\textbf{k}$, and $V$ is the volume of the cavity. 
A similar analysis of the magnetostatic modes gives 
\begin{equation}
\begin{split}
 \left < \frac{w_{sn}^{(ms)} w_{s^{'}n}^{(ms)}}{V^{ms}} \right > = \\ \int \, \frac{d \Omega_k}{2 \pi V} \, \tilde{\textbf{e}}^{*}_s \cdot 
 \left [ \left ( \frac{k^2_{\perp}}{k^2} \right ) \frac{\left ( \textbf{k} \times \hat{n} \right ) \left ( \textbf{k} \times \hat{n} \right )}{k_{\perp}^2} \right ]  \cdot \tilde{\textbf{e}}_{s^{'}} \,\, .
 \end{split}
\end{equation}
The connection between the cavity case (\ref{eqn:Zcav_multiport}) and the radiation case (\ref{eqn:Yrad_aperture}) is now apparent. 
Specifically, we note that the factors $w_{sn}^{(em)}$ are zero mean Gaussian random variables with a correlation matrix given by (\ref{eqn:corr_em_mode_ampl}). 
We can express the product $w_{sn}^{(em)} w_{s^{'}n}^{(em)}$ in terms of uncorrelated zero mean, unit width Gaussian random variables by diagonalizing the 
correlation matrix. 
We again introduce matrix notation and represent the $s s^{'}$ element of the product
\begin{equation}
\begin{split}
 \frac{w_{sn}^{(em)} w_{s^{'}n}^{(em)}}{V^{em}} = \\ 2 \sqrt{\frac{\epsilon}{\mu}} \Delta k_n 
 \left \{ \left [ \ushortdw{G}^{rad,*} \right ]^{1/2} \cdot \ushortw{w}_{n} \ushortw{w}^{T}_{n} \cdot \left [ \ushortdw{G}^{rad} \right ]^{1/2} \right \}_{s s^{'}} \,\, ,
\end{split}
\end{equation}
where $\ushortdw{G}^{rad}$ is the radiation admittance matrix (\ref{eqn:Grad_aperture}), and $\Delta k_n = \pi^2 / (k^2_n V)$ is the mean separation between 
resonant wave numbers for electromagnetic modes on a cavity of volume $V$. 
Substituting (\ref{eqn:corr_em_mode_ampl}) into (\ref{eqn:Zcav_multiport}) we have
\begin{equation}\label{eqn:Ycav_el}
 \begin{split}
  Y^{cav}_{s s^{'}} = \left \{ \sum_n \, \frac{2 i k_0 \Delta k_n}{\pi \left ( k_0^2 - k_n^2 \right )} 
  \left [ \ushortdw{G}^{rad} \right ]^{1/2} \cdot \ushortw{w}_{n} \ushortw{w}^{T}_{n} \cdot \left [ \ushortdw{G}^{rad} \right ]^{1/2} \right \}_{s s^{'}} \\
  + B^{ms}_{s s^{'}} \left ( k_0 \right ) \,\, ,
  \end{split}
\end{equation}
where in the limit of a large cavity we have approximated the sum of the pairs of Gaussian random variables representing the magnetostatic contribution to the cavity 
admittance by their average values and using $\Delta k_n = 2 \pi^2 / \left ( V k_n^2 \right )$ for magnetostatic modes converted the sum to an integral 
\begin{equation}
 \begin{split}
  \sqrt{\frac{\epsilon}{\mu}} \sum_n \frac{i}{k_0} \frac{w_{sn}^{(ms)} w_{s^{'}n}^{(ms)}}{V^{ms}} \approx  
  \sqrt{\frac{\epsilon}{\mu}} \sum_n \frac{i}{k_0} \left < \frac{w_{sn}^{(ms)} w_{s^{'}n}^{(ms)}}{V^{ms}} \right > 
  = \\ \sum_n \frac{2i}{k_0} \, \sqrt{\frac{\epsilon}{\mu}} \, \int \, \frac{k_n^2 \Delta k_n d \Omega_k}{\left ( 2 \pi \right )^3} \, \\
  \tilde{\textbf{e}}^{*}_s \cdot 
  \frac{\left ( \textbf{k}_{n \perp} \times \hat{n} \right ) \left ( \textbf{k}_{n \perp} \times \hat{n} \right )}{k_{n}^2} \cdot \tilde{\textbf{e}}_{s^{'}} 
  \approx i B^{ms}_{s s^{'}} \left ( k_0 \right ) \,\, .
 \end{split}
\end{equation}
Equation (\ref{eqn:Ycav_el}) is the random coupling model prediction for the cavity admittance. 
The last steps are to replace the exact spectrum of eigenvalues, $k_n^2$ by a spectrum produced by random matrix theory 
and to insert a loss term. We introduce a reference wave frequency and associated wave number $\omega_{ref} = k_{ref} c$, 
and we assume that the cavity is filled with a uniform dielectric with loss tangent $t_{\delta} = \epsilon_i / \epsilon_r$. 
Under these assumptions for frequencies close to the reference frequency, the frequency dependent fraction appearing in (\ref{eqn:Ycav_el}) 
can be expressed as 
\begin{equation}
 \frac{2 k_0 \Delta k}{k_0^2 - k_n^2} = \left [ \mathcal{K}_0 - \mathcal{K}_n + i \alpha \right ]^{-1} \,\, ,
\end{equation}
where
\begin{equation}
 \mathcal{K}_0 = \frac{k_0^2 - k^2_{ref}}{2 k_0 \Delta k} \approx \frac{\omega - \omega_{ref}}{\Delta \omega} \,\, ,
\end{equation}
measures the deviation in frequency from the reference frequency, and $\Delta \omega = \Delta k c$, is the mean spacing in resonant frequencies. 
The resonant wave numbers are now represented as a set of dimensionless values 
\begin{equation}
 \mathcal{K}_n = \frac{k_n^2 - k^2_{ref}}{2 k_0 \Delta k} \,\, ,
\end{equation}
which by their definition have mean spacing of unity. These are then taken to be the eigenvalues of a random matrix from the Gaussian Orthogonal Ensemble 
normalized to have mean spacing unity. Finally, the loss factor $\alpha$ is defined to be 
\begin{equation}
 \alpha = \frac{1}{2} t_{\delta} \frac{k_0}{\Delta k} \,\, .
\end{equation}
We note that the zeros of the denominator are given by $k_0 = k_n - i \alpha$, which implies $\Im \left ( \omega \right ) = - \alpha \Delta \omega = - \omega_0 / (2 Q)$. 
Thus $\alpha = \omega_0 / ( 2 Q \Delta_{\omega} )$.
This results in an expression analogous to those obtained in \cite{zao1,zao2,Antonsen_2011} for the impedance matrix
\begin{equation}
  \ushortdw{Y}^{cav} = i \,\, \textrm{Im} \left \{ \ushortdw{Y}^{rad} \right \} + \left [ \ushortdw{G}^{rad} \right ]^{1/2} \cdot
                       \ushortdw{\xi}
                       \cdot \left [ \ushortdw{G}^{rad} \right ]^{1/2} \,\, ,
\end{equation}
Thus, we have seen that in cases of ports described by planar apertures, we can express the model cavity admittance in 
terms of the corresponding radiation impedance or admittance and a universal statistical matrix $\ushortdw{\xi}$.
\bibliographystyle{IEEEtran}

\end{document}